\documentclass[aip,apl,reprint]{revtex4-1}
\usepackage{graphicx} 
\usepackage{dcolumn} 
\usepackage{bm} 
\begin{document}
\preprint{D. J. Kim \textit{et al.}}
\title{Control of defect-mediated tunneling barrier heights in ultrathin MgO films}
\author{D. J. Kim}
\email[electronic mail: ]{dong-jik.kim@ipcms.u-strasbg.fr}
\affiliation{IPCMS UMR 7504 CNRS, Universit\'e de Strasbourg, 23 Rue du Loess BP 43, 67034 Strasbourg Cedex 2, France}
\author{W. S. Choi}
\altaffiliation{current address: Materials Science and Technology Division, Oak Ridge National Lab, Oak Ridge, Tennessee 37831-6114, USA}
\affiliation{ReCFI, Department of Physics and Astronomy, Seoul National University, Seoul 151-747, Republic of Korea}
\author{F. Schleicher}
\affiliation{IPCMS UMR 7504 CNRS, Universit\'e de Strasbourg, 23 Rue du Loess BP 43, 67034 Strasbourg Cedex 2, France}
\affiliation{Institute of Physics, Wroc\l aw University of Technology, Wybrze\.{z}e Wyspia\'nskiego 27, 50-370 Wroc\l aw, Poland}
\author{R. H. Shin}
\affiliation{Department of Physics, Ewha Womans University, Seoul 120-750, Republic of Korea}
\author{S. Boukari}
\author{V. Davesne}
\author{C. Kieber}
\author{J. Arabski}
\author{G. Schmerber}
\author{E. Beaurepaire}
\affiliation{IPCMS UMR 7504 CNRS, Universit\'e de Strasbourg, 23 Rue du Loess BP 43, 67034 Strasbourg Cedex 2, France}
\author{W. Jo}
\affiliation{Department of Physics, Ewha Womans University, Seoul 120-750, Republic of Korea}
\author{M. Bowen}
\email[electronic mail: ]{martin.bowen@ipcms.u-strasbg.fr}
\affiliation{IPCMS UMR 7504 CNRS, Universit\'e de Strasbourg, 23 Rue du Loess BP 43, 67034 Strasbourg Cedex 2, France}
\date{\today}
\begin{abstract}
The impact of oxygen vacancies on local tunneling properties across rf-sputtered MgO thin films was investigated by optical absorption spectroscopy and conducting atomic force microscopy. Adding O$_2$ to the Ar plasma during MgO growth alters the oxygen defect populations, leading to improved local tunneling characteristics such as a lower density of current hotspots and a lower tunnel current amplitude. We discuss a defect-based potential landscape across ultrathin MgO barriers.
\end{abstract}
\maketitle
MgO-based tunnel barriers have been intensively studied since the theoretical prediction \cite{ButlerPRB,MathonPRB} of $>$1000\% tunneling magnetoresistance (TMR) in Fe/MgO/Fe and initial experimental confirmation.\cite{BowenMgOAPL2001,MgO100schuhlAPL2003} However, TMR ratios at room temperature in MgO-based MTJs have reached only 200\% with Fe electrodes\cite{NMAT2004Yuasa,JPSJ2008Yuasa} and 600\% with FeCoB alloy electrodes.\cite{APL2008Ikeda} The most important cause of the discrepancy is believed to be structural defects in the MgO barrier. Yet despite previous experimental\cite{PRB1990Wang,PRB1994Gibson,JCP1998Illas,PRB2003Dominguez-Ariza} and theoretical\cite{PR1950Molnar,PR1953Day,PR1969Chen,RPB1989Rosenblatt,PRB2006Mather} studies of point defects in MgO, the properties of defects in MgO films are still in debate. As technology goes nanoscale and increasingly integrates alternative dielectrics to silicon, it becomes important to understand the impact of defects on the properties of these ultrathin dielectrics such as MgO toward coherent transport\cite{LPMNancyCrTunnelPRL} and memristive effects.\cite{bow06,Halley}

To alter the density of only intrinsic MgO defects in our films,\cite{JPD2010Kim} we fixed the working pressure during rf-sputtering but modified the proportion of O$_2$ and Ar from 0\% to 10\% O$_2$. To prevent the oxidation of the Fe surface, the first half monolayer was sputtered with pure Ar gas for all samples.\cite{footnote1} To qualitatively determine the change in defect density, we measured the optical transmission of MgO(001)//MgO(50 nm) samples using a UV-Vis-NIR Perkin-Elmer Lambda 950 spectrophotometer with an integrating sphere of 150 mm. Local tunneling current maps and current-voltage (I-V) curves were obtained on MgO(001)//MgO(10 nm)/Fe(20 nm)/MgO(1.2 nm) samples grown in optimized growth conditions\cite{JPD2010Kim,thickness} using conductive atomic force microscopy (C-AFM; N-Tracer, Nanofocus Inc.). The C-AFM tip was grounded and the lower Fe electrode was biased. The spatial resolution of C-AFM is the same as that of a conventional AFM and the minimum detectable current of C-AFM is about 0.5 pA.

\begin{figure} [b]
\includegraphics{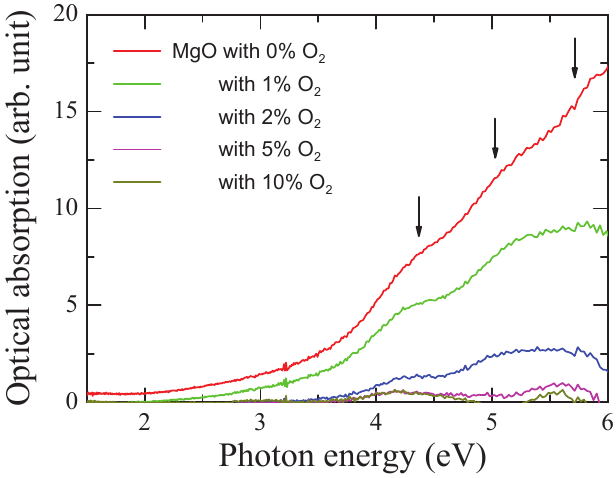}
\caption{\label{fig:figure1}(Color online) Optical absorption spectra of MgO(001)//MgO(50 nm) samples.}
\end{figure}

The spectral dependencies of the optical absorption coefficient $\alpha(\omega)$ of MgO(001)//MgO(50 nm) samples shown in Fig. 1 were extracted from experimental transmission spectra using $\alpha_{layer}(\omega) = -[ln~T_{total}(\omega)- ln~T_{sub}(\omega)]/d$, where $T(\omega)$ is the optical transmission and $d$ is the MgO layer thickness atop the MgO substrate with its own optical transmission $T_{sub}(\omega)$. Note the presence in the sample with 0\% O$_2$ of three peaks (see arrows in Fig. 1): the peak at $\sim$4.2 eV is assigned to M centers (the combination of two oxygen vacancies);\cite{PRB2003Dominguez-Ariza} the peak at $\sim$5.0 eV is assigned to F and F$^+$ centers (neutral and singly charged oxygen vacancies);\cite{JCP1998Illas,PR1969Chen} finally, a weak peak\cite{footnote2} at $\sim$5.6 eV is assigned to F$^{2+}$ center (doubly charged oxygen vacancy).\cite{PR1950Molnar,PRB1994Gibson} As the O$_2$ proportion increases, the spectral weight around each of the three peaks decreases systematically, underscoring a control during sputter-growth of oxygen defect densities in our MgO films that is more direct and effective than other methods.\cite{PR1953Day,PR1969Chen,PRB2006Mather,PRL2008Miao}

\begin{figure}
\includegraphics{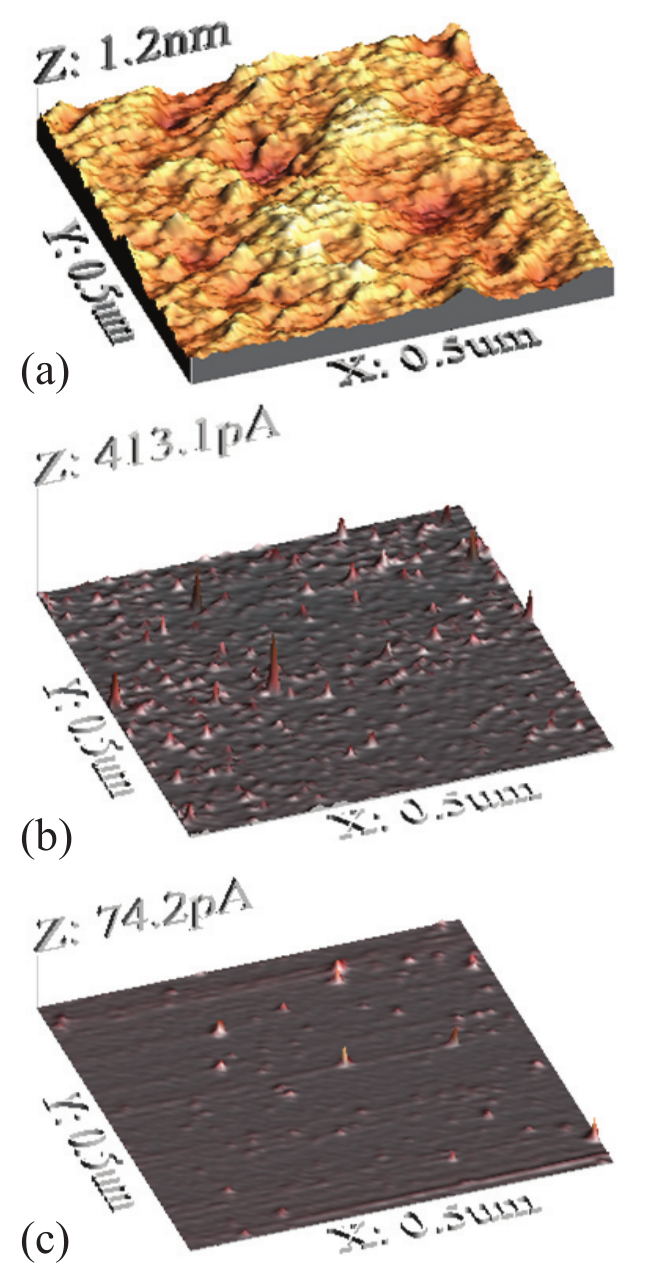}
\caption{\label{fig:figure2}(Color online) (a) Topographic map of MgO barrier in 0\%-O$_2$ MgO layer and local tunneling maps with 1 V bias of (b) 0\%- and (c) 10\%-O$_2$ MgO layers. Topographic map in 10\%-O$_2$ MgO layer is nearly the same with (a). The rms roughness of (a) is 0.14 nm.}
\end{figure}

We now examine the impact of oxygen vacancies on tunneling transport across a 1.2 nm-thick MgO layer. Figures 2(a) and 2(b) respectively show the topographic and local tunnel current maps across a 0\%-O$_2$ MgO layer with 1 V bias. The MgO layer is very flat owing to a rms roughness that is identical to that of the MgO substrate. Similar to previous results,\cite{JPD2010Kim,EPJB2000DaCosta,JMMM2009Bhutta,JAP2000Ando} we obtain an inhomogeneous local current map including a number of current hotspots. Since there is no correlation between topography and current maps, this tunnel current inhomogeneity originates from fluctuations in barrier thickness and/or height.\cite{EPL1997Bardou} However, while the topographical map of the 10\%-O$_2$ MgO layer resembles that of the 0\%-O$_2$ MgO layer (data not shown), the corresponding local tunnel current map [see Fig. 2(c)] reveals a lower density of hotspots and, moreover, a lower tunnel current amplitude. This suggests that the tunnel current inhomogeneity across MgO reflects solely a barrier height fluctuation that originates from oxygen vacancies.

\begin{figure} [b]
\includegraphics{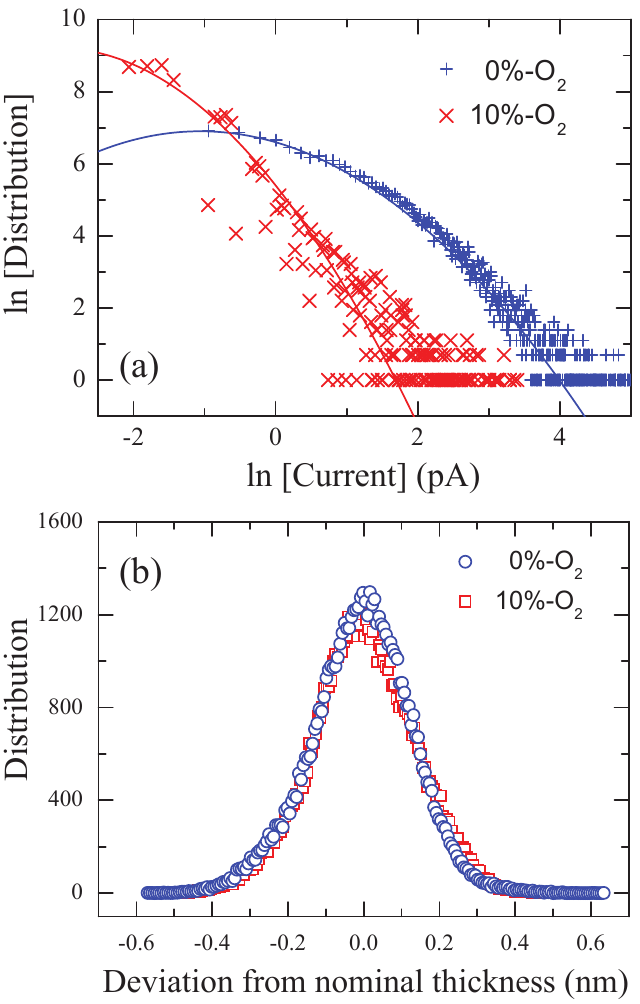}
\caption{\label{fig:figure3}(Color online) (a) Distributions of local tunnel current from Figs. 2(b) and (c). Lines are the best fits with a log-normal distribution. (b) Distributions of topographic deviation from the nominal thickness of MgO barriers in 0\%- and 10\%-O$_2$ MgO layers. These distributions fit well with a Gaussian function, where the standard deviation is 0.26 and 0.25 for 0\%- and 10\%-O$_2$ MgO layers, respectively.}
\end{figure}

In support of this suggestion, we show in Fig. 3(a) statistical distributions of local tunnel current through 0\%- and 10\%-O$_2$ MgO ultrathin films. If there is a distribution of tunneling parameters, such as barrier height/thickness, the local tunnel current is correctly described by a log-normal distribution. The tunnel current $i=i_0e^{-l/\lambda}$ depends exponentially on the barrier thickness $l$ and the attenuation length $\lambda$, where $\lambda$ is a function of barrier height. Assuming a Gaussian distribution of $l/\lambda$ with a mean $\mu$ and a standard deviation $\sigma$, the tunnel current has a log-normal distribution $P(i)$:\cite{EPJB2000DaCosta,EPL1997Bardou}
\begin{equation}
P(i)=\frac{1}{i\sqrt{2\pi\sigma^2}}{\rm{exp}}\left(-\frac{\left({\rm{ln}}\ i-\mu\right)^2}{2\sigma^2} \right).
\end{equation}
By fitting our data for 0\%- and 10\%-O$_2$ MgO layers [see lines in Fig. 3(a)], we respectively find typical currents $i_{typ}$ of 0.38 pA and 0.047 pA, averaged currents $\langle i\rangle$ of 6.1 pA and 0.30 pA, and the $\sigma$ values of $l/\lambda$ of 1.4 and 1.1. Since both MgO layers have virtually the same thickness fluctuation [see Fig. 3(b)], a lower $\sigma$ value implies a spatially more uniform barrier height, and thus the lower $i_{typ}$ and $\langle i\rangle$ imply a larger barrier height as the O$_2$ proportion is increased.

\begin{figure}
\includegraphics{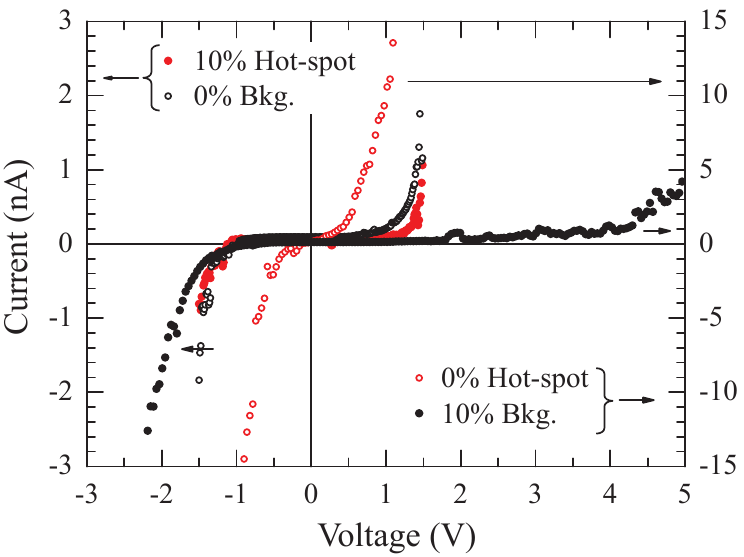}
\caption{\label{fig:figure4}(Color online) Current-voltage curves of 0\%- and 10\%-O$_2$ MgO layers, measured at hotspot and background using conductive tip.}
\end{figure}

To understand this increased barrier height from a defect perspective, we present in Fig. 4 typical local I-V curves taken away from and on current hotspots.\cite{footnote3} Since the first MgO half monolayer was systematically grown in a pure Ar gas to avoid Fe oxidation, the I-Vs for the 10\%-O$_2$ MgO may be asymmetric, with data at negative bias (thus probing the bottom interface) that resembles that for a 0\%-O$_2$ MgO sample. We therefore focus only on data acquired at positive bias, corresponding to the top interface, and qualitatively pinpoint the barrier heights. Going from 0\% to 10\% O$_2$ raises the barrier height from 0.9 eV to 2.3 $\sim$ 4.1 eV in the background,\cite{footnote4} and from 0.4 eV to 1.1 eV on hotspots. Interestingly, the I-Vs for 0\%-O$_2$/`off-hotspot' and 10\%-O$_2$/`on-hotspot' are quite similar. This suggests that going from 0\% to 10\% O$_2$ eliminates hotspot-inducing defects responsible for the 0.4 eV barrier height in the 0\%-O$_2$ sample, while the defects responsible for the 0.9 eV background barrier in the 0\%-O$_2$ sample now account in the 10\%-O$_2$ sample only for hotspots relative to an improved background. In fact, for a given positive bias, the tunneling current is lower for 10\%-O$_2$/`on-hotspot' than for 0\%-O$_2$/`off-hotspot', implying that the density of the defect type(s) responsible for this I-V characteristic has decreased. Naively, if we assume Fermi level pinning in the middle of the bulk MgO band gap\cite{BowenPRB2006,footnote5} of 7.8 eV and the presence at 5 eV above the top of the valence band of F/F$^+$ defect states,\cite{PRB1994Gibson} then F/F$^+$ defects define an effective barrier height of $\sim$1.0 eV that might correspond to that observed in the I-V data for the 0\%-O$_2$ background and 10\%-O$_2$ hotspot data, consistently with the near extinction of F/F$^+$ absorption at 5 eV as the oxygen proportion is increased (see Fig. 1). A more quantitative description of the defect-mediated potential landscape across ultrathin MgO would require a) mitigating the impact of water exposure between the growth and characterization phases of the study; b) determining the evolution of the band gap across the atomic planes of the ultrathin film so as to account for metal-induced gap states\cite{BowenPRB2006} and size effects;\cite{LTOSTOEELS} and c) considering the likely surface/interface nature of defects in the ultrathin film, whose transition energies and positions within the band gap are different from those of the bulk and remain largely unstudied.\cite{PRB2003Dominguez-Ariza}

In conclusion, we have shown how to control the nature and density of oxygen defects during the growth of rf-sputtered MgO thin and ultrathin films via the combination of optical absorption and local tunneling current maps obtained by conductive atomic force microscopy. Eliminating certain defect populations leads to improved barrier heights with a reduced spatial fluctuation, and to a change in the type of defect-induced hotspot that drives tunneling transport in a micronic device.

We thank Y. Yin for assistance with optical experiments, and D. Halley and S. J. Park for stimulating discussions. This research was supported by the EC FP6 (NMP3-CT-2006-033370), the French ANR (ANR-06-NANO-033-01, ANR-09-JCJC-0137), and by the Korean MEST (2010K000339 from CNMT under 21st Century Frontier R\&D Programs; NRF 2010-0020416).


\begin{thebibliography}{19}
\bibitem{ButlerPRB} W. H. Butler, X.-G. Zhang, T. C. Schulthess, and J. M. Maclaren Phys. Rev. B \textbf{63}, 054416 (2001).
\bibitem{MathonPRB} J. Mathon and A. Umerski, Phys. Rev. B \textbf{63}, 220403R (2001).
\bibitem{BowenMgOAPL2001} M. Bowen \textit{et al.}, Appl. Phys. Lett. \textbf{79}, 1655 (2001).
\bibitem{MgO100schuhlAPL2003} J. Faure-Vincent \textit{et al.}, Appl. Phys. Lett. \textbf{82}, 4507 (2003).
\bibitem{NMAT2004Yuasa} S. Yuasa, T. Nagahama, A. Fukushima, Y. Suzuki, and K. Ando, Nature Mater. \textbf{3}, 868 (2004).
\bibitem{JPSJ2008Yuasa} S. Yuasa, J. Phys. Soc. Jpn. \textbf{77}, 031001 (2008).
\bibitem{APL2008Ikeda} S. Ikeda \textit{et al.}, Appl. Phys. Lett. \textbf{93}, 082508 (2008).
\bibitem{PRB1990Wang}
Q. S. Wang and N. A. Holzwarth, Phys. Rev. B \textbf{41}, 3211 (1990).
\bibitem{PRB1994Gibson}
A. Gibson, R. Haydock, and J. P. LaFemina, Phys. Rev. B \textbf{50}, 2582 (1994).
\bibitem{JCP1998Illas}
F. Illas and G. Pacchioni, J. Chem. Phys. \textbf{108}, 7835 (1998).
\bibitem{PRB2003Dominguez-Ariza}
D. Domingues-Ariza, C. Sousa, F. Illas, D. Ricci, and G. Pacchioni, Phys. Rev. B \textbf{68}, 054101 (2003).
\bibitem{PR1950Molnar} J. P. Molnar and C. D. Hartman, Phys. Rev. \textbf{79}, 1015 (1950).
\bibitem{PR1953Day}
H. R. Day, Phys. Rev. \textbf{91}, 822 (1953).
\bibitem{PR1969Chen}
Y. Chen, T. Williams, and W. A. Sibley, Phys. Rev. \textbf{182}, 960 (1969).
\bibitem{PRB2006Mather}
P. G. Mather, J. C. Read, and R. A. Buhrman, Phys. Rev. B \textbf{73}, 205412 (2006).
\bibitem{RPB1989Rosenblatt}
G. H. Rosenblatt, M. W. Rowe, G. P. Williams, Jr., R. T. Williams, and Y. Chen, Phys. Rev. B \textbf{39}, 10309 (1989).
\bibitem{LPMNancyCrTunnelPRL} F. Greullet \textit{et al.}, Phys. Rev. Lett. {\bf 99}, 187202, (2007).
\bibitem{bow06} M. Bowen \textit{et al.}, Appl. Phys. Lett. \textbf{89}, 103517 (2006).
\bibitem{Halley}
D. Halley \textit{et al.}, Appl. Phys. Lett. {\bf 92}, 212115 (2008).
\bibitem{footnote1}
Any Fe oxidation would lead to states near the middle of the MgO band gap, and thus to a lower barrier height in contrast to our results. See P. A. Cox, Transition Metal Oxides, Oxford Science Publications, Oxford, Chap. 2 (1992).
\bibitem{JPD2010Kim}
D. J. Kim \textit{et al.}, J. Phys. D: Appl. Phys. \textbf{43}, 215003 (2010).
\bibitem{thickness}
Careful thickness calibrations using a combination of \textit{ex-situ} stylus profilometry and an \textit{in-situ} quartz microbalance enable an effective relative statistical error of 0.1 nm on our 1.2 nm-thick films.
\bibitem{footnote2}
Low resolution reflects here a loss of transmission across MgO.
\bibitem{PRL2008Miao}
G. X. Miao, Y. J. Park, J. S. Moodera, M. Seibt, G. Eilers, and M. Munzenberg, Phys. Rev. Lett. \textbf{100}, 246803 (2008).
\bibitem{EPJB2000DaCosta}
V. Da Costa, Y. Henry, F. Bardou, M. Romeo, and K. Ounadjela, Eur. Phys. J. B \textbf{13}, 297 (2000).
\bibitem{JMMM2009Bhutta}
K. M. Bhutta, J. Schmlhorst, and G. Reiss, J. Magn. Magn. Mater. \textbf{321}, 3384 (2009).
\bibitem{JAP2000Ando}
Y. Ando, H. Kameda, H. Kubota, and T.Miyazaki, J. Appl. Phys \textbf{87}, 5206 (2000).
\bibitem{EPL1997Bardou}
F. Bardou, Europhys. Lett. \textbf{39}, 239 (1997).
\bibitem{footnote3}
Nonlinear data excludes direct conduction between the tip and the lower Fe layer even on the hotspot.
\bibitem{footnote4}
Depending on the local background spot considered, current increases may appear at 2.3, 3.1, 3.6, and 4.1 eV.
\bibitem{BowenPRB2006}
M. Bowen \textit{et al.}, Phys. Rev. B {\bf 73}, 140408R (2006).
\bibitem{footnote5}
This is consistent with barrier heights of 3.6 eV and 3.9 eV found for MgO/Fe and MgO/Au. See W. Wulfhekel \textit{et al.}, Appl. Phys. Lett. \textbf{78}, 509 (2001) \& S. Gu\'ezo \textit{et al.}, Appl. Phys. Lett. \textbf{93}, 172116 (2008).
\bibitem{LTOSTOEELS}
A. Ohtomo, D. A. M\"{u}ller, J. L. Grazul, and H. Y. Hwang, Nature \textbf{419}, 378 (2002).
\end{thebibliography}
\end{document}